\documentclass[12pt]{iopart}

\usepackage{graphicx}
\usepackage{color}

\begin{document}

\title[Dominance of many-body effects over one-electron mechanism in Nd$_{2-x}$Ce$_x$CuO$_4$]
{Dominance of many-body effects over one-electron mechanism for band structure
doping dependence in Nd$_{2-x}$Ce$_x$CuO$_4$: LDA+GTB approach}

\author{M M Korshunov$^{1,2}$, V A Gavrichkov$^{1}$, S G Ovchinnikov$^{1}$,
I A Nekrasov$^{3}$, E E Kokorina$^{3}$, Z V Pchelkina$^{4}$}

\address{$^{1}$ L.V. Kirensky Institute of Physics, Siberian Branch of Russian Academy of Sciences, 660036 Krasnoyarsk, Russia}
\address{$^{2}$ Max-Planck-Institut f\"{u}r Physik komplexer Systeme, D-01187 Dresden, Germany}
\address{$^{3}$ Institute of Electrophysics, Russian Academy of Sciences-Ural Division, 620016 Yekaterinburg, Amundsena 106, Russia}
\address{$^{4}$ Institute of Metal Physics, Russian Academy of Sciences-Ural Division, 620041 Yekaterinburg, GSP-170, Russia}
\ead{maxim@mpipks-dresden.mpg.de}

\begin{abstract}
In the present work we report the band structure calculations for
the high temperature superconductor Nd$_{2-x}$Ce$_x$CuO$_4$ in the
regime of strong electronic correlations within an LDA+GTB method,
which combines the local density approximation (LDA) and the generalized
tight-binding method (GTB).
The two mechanisms of band structure doping dependence were taken into account. Namely, the one-electron mechanism provided by the doping dependence of the crystal structure, and the many-body mechanism provided by the strong renormalization of the fermionic quasiparticles due to the large on-site Coulomb repulsion.
We have shown that in the antiferromagnetic and in the strongly correlated paramagnetic phases of the underdoped cuprates the main contribution to the doping evolution of band structure and Fermi surface comes from the
many-body mechanism.
\end{abstract}

\submitto{\JPCM}
\maketitle

\section{Introduction}

It is well known that the hole doping of La$_{2-x}$Sr$_x$CuO$_4$
and the electron doping of Nd$_{2-x}$Ce$_x$CuO$_4$ shift them
into the superconducting state. With the increase of doping concentration
$x$ the band structure undergoes dramatic changes from an antiferromagnetic (AFM) insulator to a normal paramagnetic metal. It is well established
that the strong electronic correlations play one of the main roles
in the formation of the electronic structure of High-T$_c$ cuprates.
They are important especially for small $x$ values and should be
taken into account in the calculations explicitly.

In the present paper to study the band structure of Nd$_{2-x}$Ce$_x$CuO$_4$ we
apply the LDA+GTB computational scheme \cite{mmk2005}. This scheme combines
{\it ab-initio} calculations within the Local Density Approximation (LDA)
and the many-body Generalized Tight-Binding (GTB) approach \cite{vag2000}.
Earlier, within the GTB method a specific many-body mechanism of band structure evolution for the doped Mott-Hubbard insulator was found.
This mechanism is caused by the changes of occupation factors of many-body
electronic terms by the quasiparticle excitations \cite{vag2000}.
For Nd$_{2-x}$Ce$_x$CuO$_4$ such terms are $d^{10}p^6$ term (zero holes per unit cell) and hybridized $d^9p^6+d^{10}p^5$ term (one hole per unit cell). Here we highlight also the so-called ``one-electron'' mechanism of band structure doping dependence. It originates from the changes of a crystal structure (lattice parameters, atomic positions) upon doping. As a result, matrix elements of $d-d$, $p-d$, and $p-p$ hybridization (hoppings, and all the rest depending on the interatomic distance) vary with doping. In case of absence of a strong electronic correlations this mechanism will be responsible for evolution of band structure with doping in the framework of the standard band theory. That is why we call it one-electron mechanism.

Previously, the band structure of Nd$_{2-x}$Ce$_x$CuO$_4$ was considered
within the LDA calculations \cite{Massidda1989,Yu1991,Yu1991_2} and the tight-binding approach \cite{Mishonov2000}. Note, our approach is significantly different from these two and their simple combination. From the LDA band structure we extract the doping-dependent tight-binding parameters using projection procedure, not fitting. Then, we use a {\it many-body} tight-binding computational scheme called GTB approach. One should not mix up the one-electron tight-binding approach and many-body GTB method. The details of latter will be given in the next Section.

To our knowledge there is no band structure calculations for the High-T$_c$ cuprates that take both one-electron and many-body mechanisms into account. In the present paper we report the results of such calculations for Nd$_{2-x}$Ce$_x$CuO$_4$ and compare them with the simplified calculation which do not contain the one-electron mechanism. It was found that in the antiferromagnetic phase for small doping concentrations, $x \leq 0.1$, the  one-electron mechanism results in a slight shift of the bottom of the conduction band and, simultaneously, in a small shift of the top of the valence band, thus retaining the value of the charge-transfer gap. For higher doping concentrations the paramagnetic spin-liquid phase was considered in an approximation taking the static spin-spin correlation functions into account. In this phase, the one-electron mechanism provides small changes to the bandwidth. However, the Fermi surface and critical concentrations at which the Fermi surface topology changes, remain the same as in the case when the one-electron mechanism is disregarded.

\section{Brief description of a hybrid LDA+GTB computational scheme}

{\it Ab-initio} electronic structure calculations within the density
functional theory have their development within the LDA approximation.
This approximation does not take strong electronic correlations into account
properly. That is why the true band structure of Mott insulators can not be
described within LDA.
We employ the LDA to calculate the non-interacting part of the multi-band $p-d$ model Hamiltonian. Then, the strong electronic correlations enter in the framework of the GTB approach \cite{mmk2005}.

LDA+GTB method consists of the following steps:
\begin{enumerate}
	\item {\it Ab-initio} LDA band structure calculation, finding of Bloch functions;
	\item Construction of the Wannier functions for the physically relevant states;
	\item Construction of the multi-band $p-d$ model Hamiltonian with the parameters obtained from the previous two steps;
	\item Splitting of the multi-band $p-d$ model Hamiltonian into a sum of inter- and intra-cell components and exact diagonalization of the intra-cell part in order to construct the many-body molecular orbitals for the unit cell;
	\item Construction of the intra-cell Hubbard $X$-operators on the basis of these molecular orbitals and rewriting the full Hamiltonian for the crystal in the $X$-representation;
	\item Calculation of a quasiparticle band structure for Hubbard fermions in the framework of the perturbation theory with the small inter-cell hopping and interactions.
\end{enumerate}

Note, the first version of the GTB method \cite{vag2000,vag2004} contained a large number of unknown model parameters, which were extracted by the comparison with experimental data. In the generalized LDA+GTB method, all parameters of the theory are calculated explicitly. For a given doping concentration $x$ we calculate the {\it ab-initio} band structure. One-electron mechanism of band structure doping dependence is determined by the dependence of the matrix elements of the inter-atomic hopping and one-electron energies on doping, which is due to the change of the lattice parameters with $x$. Many-body mechanism arises from the doping dependence of the occupation factors. Thus, one-electron mechanism takes place for the ordinary tight-binding method while the many-body mechanism appears within the GTB method as the effect of the strong electronic correlations.

\section{Electronic structure and model parameters of Nd$_{2-x}$Ce$_x$CuO$_4$: LDA results \label{section:LDA}}

Nd$_2$CuO$_4$ crystallizes in tetragonal structure with the symmetry space
group I4/mmm \cite{tk1994}, also called the T'-structure.
Lattice parameters are $a=b=3.94362$\AA, and $c=12.1584$ \AA \cite{tk1994}.
Cu ions in positions $2a$, $(0, 0, 0)$, are surrounded by four ions of oxygen O1, which occupy positions $4c$, $(0, 0.5, 0)$. Ions of Nd in positions
$4e$, $(0, 0, 0.35112)$, have eight nearest neighbors of oxygen O2 in
positions $4d$, $(0, 0.5, 0.25)$. In comparison with the high-temperature
tetragonal structure of La$_2$CuO$_4$, in the T'-structure of Nd$_2$CuO$_4$
the apical oxygen ions around Cu ions are absent. With Ce doping
the symmetry group of Nd$_{2-x}$Ce$_x$CuO$_4$ remains the same whereas
the lattice parameters and $z$-coordinate of Nd positions are
changing (see Table~\ref{table1}) \cite{tk1994,efp1990}.

\begin{figure}
\centering
\includegraphics[width=0.8\linewidth]{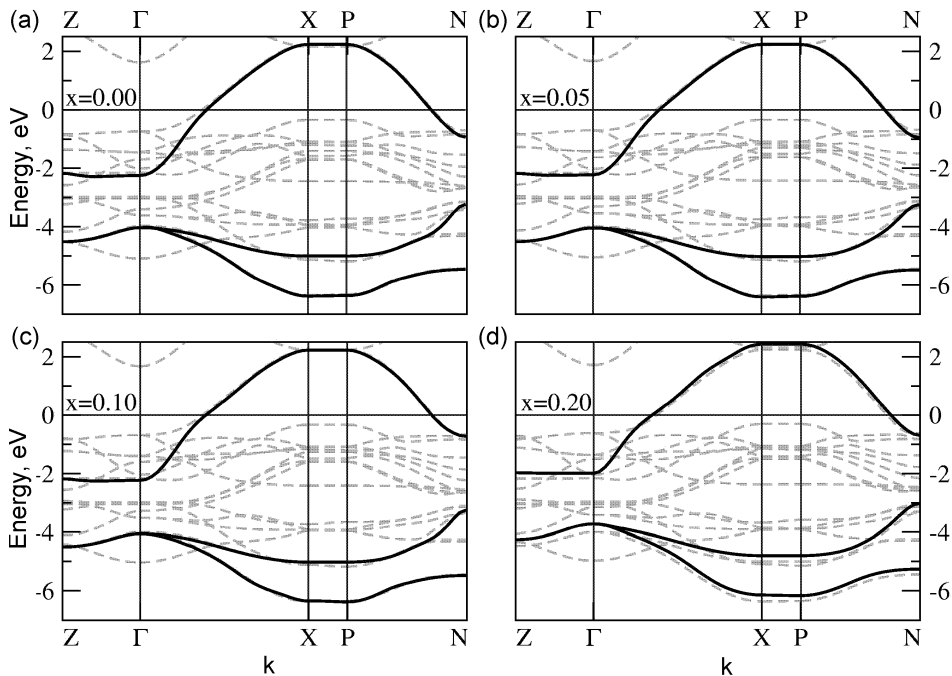}
\caption{\label{fig2} The comparison of the electronic dispersion obtained
within the LDA calculations (gray dashed curves) and the electronic dispersion of the effective non-interacting 3-band Hamiltonian for the NMTO orbitals basis set (black solid curves) for different Ce concentrations $x$.}
\end{figure}

Electronic structure of Nd$_{2-x}$Ce$_x$CuO$_4$ at $0 \leq x \leq 0.3$ was
calculated using the local density approximation. To this end,
the linearized muffin-tin orbitals (LMTO) method in the tight-binding
approach within atomic spheres approximation (TB-LMTO-ASA) was applied
\cite{oka1984,oka1986}. The $4f$ states of Nd were considered as semi-core states, because they are well localized and are situated far below the Cu-$d$ states \cite{djs1991}. The LDA band structure of Nd$_{2-x}$Ce$_x$CuO$_4$ along the high symmetry directions of the Brillouin zone is shown in Fig.~\ref{fig2} with gray dashed curves. Bands formed by the hybridized $3d$ copper and $2p$ oxygen states has a width of approximately 9 eV. As a result of hybridization between $d_{x^2-y^2}$ copper orbital and appropriate $p_{x,y}$ orbitals of plain oxygen (O1) the bonding bands are located at energies $-5...-6$ eV, while the antibonding bands crosse the Fermi level. These hybrid orbitals determine the non-interacting Hamiltonian of the so-called 3-band $p-d$ model \cite{emery1,varma1}.

\begin{table}
\caption {\label{table1} Crystal structure parameters (\AA) for Nd$_{2-x}$Ce$_x$CuO$_4$ for different Ce concentrations (see Refs.~\cite{tk1994,efp1990}).}
\begin{indented}
\item[] \begin{tabular}{lrrrrrr}
\br
Lattice constant&	$x=$ 0.00&	0.05&	0.10&	0.15&	0.20&	0.30\\
\mr
$a$&	3.94362&	3.94056&	3.94071&	3.94224&	3.94295&	3.94288\\
$c$&	12.1584&	12.1130&	12.0945&	12.0603&	12.030&	12.0288\\
$z$(Nd)&	0.35112&	0.3519&	0.3523&	0.3527&	0.3531&	0.3531\\
\br
\end{tabular}
\end{indented}
\end{table}

To calculate hopping integrals for the 3-band $p-d$ model the NMTO
(muffin-tin orbitals of N-th order) \cite{oka2000} method was used.
For the three physically relevant hybrid bands (Cu-$d_{x^2-y^2}$ and O1-$p_{x,y}$) the NMTO orbitals were constructed.
Corresponding band dispersions are shown in
Fig.~\ref{fig2} by solid black curves. Observe, that NMTO orbitals
are almost perfectly reproduce the LDA calculated bonding and
antibonding bands formed by a hybrid $d_{x^2-y^2}$ copper and
$p_{x,y}$ oxygen orbitals. Thus, within this NMTO basis we have
obtained an effective few-orbital Hamiltonian, the 3-band $p-d$ model,
which we were looking for.

Note, we do not take into account the Cu-$4s$ orbitals explicitly.
Their importance for the bilayer cuprates was shown in Refs. \cite{Andersen1995,Andersen1996,Pavarini2001}. For bilayer structures, the $4s$ states connect two neighboring CuO$_2$-layers. It is not the case for Nd$_{2-x}$Ce$_x$CuO$_4$, the system where there is only one CuO$_2$-layer per unit cell. Moreover, due to the absence of apical oxygens this layer becomes even more two-dimensional.
There is also a contribution from the Cu-$4s$ states to the in-plane hopping integrals \cite{Mishonov2000}. However, in paper \cite{Andersen1995} it was shown that one can do the so-called Loewdin transformation and end up with the 3-band $p-d$ model with the renormalized parameters. Since our tight-binding parameters are not fitted but calculated within the NMTO method from the full-band LDA results, such renormalization is automatically performed. Thus, we partially take into account the Cu-$4s$ states implicitly.

\begin{table}
\caption {\label{table2} Hopping integrals and one-electron
energies (eV) as a function of Ce concentration for Nd$_{2-x}$Ce$_x$CuO$_4$
obtained by the NMTO method. Here $x^2$, $p_x$, $p_y$ denote the
Cu-$d_{x^2-y^2}$ and O1-$p_{x,y}$ orbitals respectively.}
\begin{indented}
\item[] \begin{tabular}{lcrrrrrr}
\br
&&	$x=$ 0.00&	0.05&	0.10&	0.15&	0.20&	0.30\\
\mr
Energy&&&&&&&\\
\mr
$E(x^2)$&	&	-2.2855&	-2.2847&	-2.1760&	-2.4215&	-2.3507&	 -2.3234\\
$E(p_x)$&	&	-3.2935&	-3.3064&	-3.2829&	-3.2607&	-3.2800&	 -3.2957\\
\mr
Hopping&	Direction&&&&&&\\
\mr
$t(x^2,p_x)$&	(0.5, 0)&	1.1216&	1.1454&	1.1665&	1.1614&	1.1726&	1.1719\\
$t'(x^2,p_x)$&	(0.5, 1)&	-0.0504&	-0.0359&	-0.0211&	-0.0202&	 -0.0166&	-0.0201\\
$t''(x^2,p_x)$&	(1.5, 0)&	0.0834&	0.0921&	0.1173&	0.1130&	0.1203&	0.1126\\
$t'''(x^2,p_x)$&	(1.5, 1)&	-0.0149&	-0.0083&	0.0015&	0.0090&	0.0153&	 0.0099\\
$t(p_x,p_y)$&	(0.5, 0.5)&	0.8320&	0.8389&	0.8381&	0.8365&	0.8386&	0.8304\\
$t'(p_x,p_y)$&	(1.5, 0.5)&	0.0266&	0.0331&	0.0452&	0.0450&	0.0469&	0.0388\\
\br
\end{tabular}
\end{indented}
\end{table}

Having applied the Fourier transform to this effective Hamiltonian
in momentum space we obtain the real space hopping integrals depending
on the distances between atoms (see Table~\ref{table2}). From Fig.~\ref{fig2}
and Table~\ref{table2} one can conclude that hopping integrals change
with doping slightly. It allows us to assume that the one-electron
contribution to the evolution of Nd$_{2-x}$Ce$_x$CuO$_4$ electronic
structure with increase of Ce concentration is not substantial.

For the 3-band $p-d$ model we also need the values of Coulomb repulsion $U$ and Hund's exchange parameter $J_H$ for Cu ions. In the paper \cite{via2002} they
were calculated for copper in the La$_2$CuO$_4$ compound by a super-cell
method \cite{og1989} and equal to: $U=10$ eV, $J_H=1$ eV. We presume these
values to be doping independent and use them in the present paper
for the Nd$_{2-x}$Ce$_x$CuO$_4$ compound.

\section{LDA+GTB results in the AFM phase}

For the wide range of doping concentrations Nd$_{2-x}$Ce$_x$CuO$_4$ remains
in the AFM phase. Therefore, we will consider the evolution
of the band structure with doping in the AFM phase first.
Within the GTB method for this phase we use the Hubbard-I approximation
\cite{hubbard1963}, though the diagram technique for the $X$-operators
\cite{zaitsev1975, izumov1991, ovchinnikov_book2004} allows one to go
beyond this approximation.

To write down the model Hamiltonian we use the Hubbard $X$-operators \cite{hubbard1964}:
$X_f^\alpha \leftrightarrow X_f^{n,n'} \equiv \left| n \right> \left< n' \right|$.
Here index $\alpha \leftrightarrow (n,n')$ enumerates quasiparticles with energy $\omega_\alpha = \varepsilon_n (N + 1) - \varepsilon_{n'} (N)$,
where $\varepsilon_n$ is the $n$-th energy level of the $N$-electron system.
The commutation relations between $X$-operators are quite complicated,
i.e. two operators commute on another operator, not a $c$-number.
Nevertheless, depending on the difference of the number of fermions in states $n$ and $n'$
it is possible to define a quasi-Fermi and a quasi-Bose type operators in terms of obeyed statistics.
There is a simple correspondence between $X$-operators and the single-electron
annihilation operators:
$a_{f \lambda \sigma} = \sum\limits_\alpha \gamma_{\lambda \sigma}(\alpha) X_f^\alpha$,
where the coefficients $\gamma_{\lambda \sigma}(\alpha)$ determine the partial weight of the quasiparticle $\alpha$ with spin $\sigma$ and orbital index $\lambda$. These coefficients are calculated straightforwardly within the GTB scheme.

In the Hubbard-I approximation the dispersion equation for the band
structure of the Hubbard fermions in the AFM phase with the sublattices
$A$ and $B$ is the following \cite{vag2004}:
\begin{equation}
\label{eqGTBdisp}
\left\| \left( E - \Omega^B_\alpha \right) \delta_{\alpha \beta} / F_{\alpha}^B - 2 \sum\limits_{\lambda \lambda'}
{\gamma_{\lambda \sigma}^*(\alpha) T_{\lambda \lambda'}^{AB}({\vec k}) \gamma_{\lambda' \sigma}(\beta)} \right\| = 0,
\end{equation}
where $\Omega^B_\alpha$ is the intra-cell local energy of the Hubbard's
fermion, $T_{\lambda \lambda'}^{AB}({\vec k})$ is the Fourier transform of
the matrix element of the intra-cell hopping between the one-electron orbitals
$\lambda$ and $\lambda'$. The occupation factor, $F_{\alpha}^B$,
is equal to the sum of the occupancies of initial and final many-body states, the transition between which is described by the operator $X_f^\alpha$.

\begin{figure}
\centering
\includegraphics[width=0.8\linewidth]{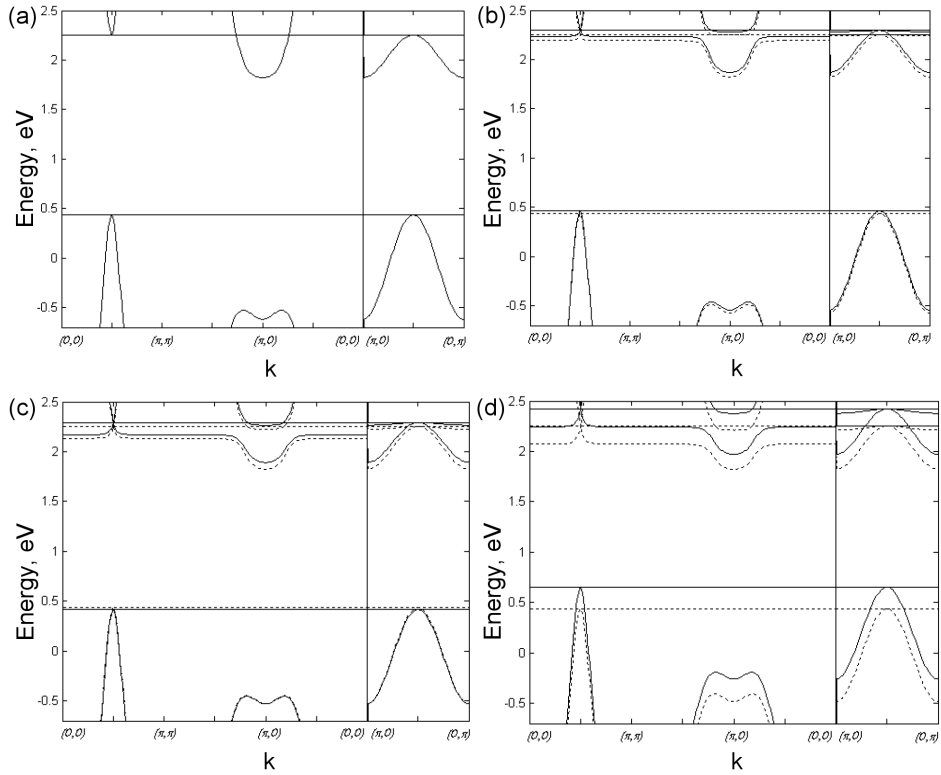}
\caption{\label{fig3} Quasiparticle's energy momentum dependence
calculated within the LDA+GTB method for $x=0$ (a), $x=0.05$ (b), $x=0.1$ (c), and $x=0.15$ (d). The dotted curve corresponds to the calculation without
one-electron mechanism of concentration dependence. The solid curve represents results of calculations in which both many-body and one-electron mechanisms were considered.}
\end{figure}

Since the structure of dispersion equation (\ref{eqGTBdisp}) is similar
to the equation in the one-electron tight-binding approach, our method
was called Generalized Tight-Binding (GTB) method. At the same time,
there are important distinctions of equation (\ref{eqGTBdisp}) from
its one-electron analogue. Namely, the local energies $\Omega^B_\alpha$ are
calculated with explicit consideration of strong electronic correlations
inside the cell. Thus, corresponding occupation factor can have non-integer
values and it depends on temperature and doping concentration.
As a result, the doping dependence of the electronic structure is not
described by the rigid band model, so the effect of doping is not only
due to the shift of a chemical potential for a given band structure.
There are bands with the spectral weight proportional to $x$ for small doping concentrations. These bands appear inside the dielectric gap near of the bottom of the conduction band for n-type cuprates and near of the top of the valence band in hole-doped cuprates.
These states are called ``in-gap states'' \cite{vag2000,vag2004,sgo2004}.
Note, they are not an impurity states of a doped semiconductor
formed in the presence of a defect since no defects are present in our model.
An example of the in-gap states near of the bottom of the conduction band in
Nd$_{2-x}$Ce$_x$CuO$_4$ will be discussed later in this section.

The results of the LDA+GTB calculations of the dispersion using equation (\ref{eqGTBdisp}) in the Neel state are shown in Fig.~\ref{fig3}.
In each figure for a given doping concentration the results of two calculations are shown:
(i) with the model parameters calculated for $x=0$ (i.e. in absence of one-electron mechanism but with the many-body mechanism taken into account), and
(ii) for the model parameters which depend on the concentration (see Table~\ref{table2}), also taking the many-body mechanism into account.
In the former case, the doping dependence is determined only by the doping-dependence of the occupation factors, $F_{\alpha}^B$, while in the latter case the dependence of matrix elements $T_{\lambda \lambda'}^{AB}({\vec k})$ on doping was also taken into account.

The results of the electronic structure calculations for an undoped
compound, Fig.~\ref{fig3}(a), reproduce the main effects of the strong electronic correlations in this material. On the bottom of the conduction band and on the top of the valence band there are in-gap states with the spectral weight proportional to the doping level \cite{vag2000,vag2004,sgo2004}. Upon increase of doping concentration the in-gap state at the bottom of the conduction band becomes dispersive with non-zero spectral weight [see Fig.~\ref{fig3}(b-d)].

For each concentration one can notice that ``switching-off'' the one-electron mechanism leads to a shift of the top of the valence band and the bottom of the conduction band. This shift is almost uniform and its value is very small. Also, this does not prevent appearance of the in-gap states.
Thus, one can conclude that such fine tuning of the Hamiltonian parameters gives just a shift of the electronic structure as a whole in the vicinity of the dielectric gap. This is most probably not very important for the physics of the High-T$_c$ materials.

\section{Paramagnetic phase}

To treat a spin-liquid phase, the multiband $p-d$ model Hamiltonian
was mapped onto an effective low-energy model \cite{mmk2005}.
Parameters of this effective model are obtained directly from the
{\it ab-initio} parameters of the multiband model.
The low-energy model for n-type cuprates is the $t-t'-t''-J^*$ model
($t-t'-t''-J$ model with the three-site correlated hoppings)
with the following Hamiltonian in the hole representation:
\begin{equation}
\label{eq:H}
H=\sum_{f, \sigma} \left( \varepsilon_0 -\mu \right) X_{f}^{\sigma, \sigma} +
\sum_{f \neq g, \sigma} t_{fg} X_{f}^{\sigma, 0} X_{g}^{0, \sigma}
+ \sum_{f \neq g} J_{fg} \left( \vec{S}_f \vec{S}_g - \frac{1}{4} n_f n_g \right) + H_3.
\end{equation}
Here $\mu$ is the chemical potential, $\vec{S}_f$ is the spin operator,
$S_f^+=X_f^{\sigma, \bar \sigma}$, $S_f^-=X_f^{\bar \sigma, \sigma}$,
$S_f^z=\frac{1}{2} \left( X_f^{\sigma, \sigma}-X_f^{\bar \sigma, \bar \sigma} \right)$,
$n_f=\sum \limits_{\sigma} X_f^{\sigma, \sigma}$ is the number of particles operator,
$J_{fg} = 2 \tilde{t}_{fg}^{2} / E_{ct}$ is the exchange parameter,
$E_{ct}=2$ eV is the charge-transfer gap.
In the notations of Ref.~\cite{mmk2005} the hopping matrix elements $t_{fg}$
corresponds to $-t_{fg}^{00}$, and $\tilde{t}_{fg} = -t_{fg}^{0S}$.
Hamiltonian $H_3$ contains the three-site interaction terms:
\begin{equation}
\label{eq:H_3}
H_3 = \sum\limits_{f \neq g \neq m, \sigma} \frac{\tilde{t}_{fm} \tilde{t}_{mg}}{E_{ct}}
\left( X_f^{\sigma 0} X_m^{\bar \sigma \sigma} X_g^{0\bar \sigma}
- X_f^{\sigma 0} X_m^{\bar \sigma \bar \sigma} X_g^{0\sigma} \right).
\end{equation}

In the considered case there is only one Fermi-type quasiparticle, $\alpha=(0,\sigma)$,
with $\gamma_{\lambda \sigma}(\alpha)=1$,
and the Hamiltonian in the general form in the momentum representation is given by:
\begin{equation}
H=\sum\limits_{\vec{k}, \sigma} \left( \varepsilon_0 - \mu \right) X^{\sigma,\sigma}_{\vec{k}}
+ \sum\limits_{\vec{k}} \sum\limits_{\alpha,\beta} t_{\vec{k}}^{\alpha \beta} {X_{\vec{k}}^\alpha}^\dag X_{\vec{k}}^{\beta}
+ \sum\limits_{\vec{p}, \vec{q}} \sum\limits_{\alpha,\beta,\sigma,\sigma'} V_{\vec{p} \vec{q}}^{\alpha \beta, \sigma \sigma'} {X_{\vec{p}}^\alpha}^\dag X_{\vec{p}-\vec{q}}^{\sigma,\sigma'} X_{\vec{q}}^{\beta}.
\end{equation}

The Fourier transform of the two-time retarded Green function in the energy representation,
$G_{\lambda}(\vec{k},E) = \left<\left< a_{\vec{k} \lambda \sigma} \left| a_{\vec{k} \lambda \sigma}^\dag \right. \right>\right>_E$,
can be rewritten in terms of the matrix Green function,
$\left[ \hat{D}(\vec{k},E) \right]_{\alpha \beta} = \left<\left< X_{\vec{k}}^\alpha \left| {X_{\vec{k}}^\beta}^\dag \right. \right>\right>_E$:
\begin{equation}
G_{\lambda}(\vec{k},E) = \sum\limits_{\alpha,\beta} \gamma_{\lambda \sigma}(\alpha) \gamma_{\lambda \sigma}^*(\beta)
D^{\alpha \beta}({\vec{k}},E).
\end{equation}

The generalized Dyson equation for the Hubbard $X$-operators \cite{ovchinnikov_book2004} in the paramagnetic phase ($\left< X_{0}^{\sigma,\sigma}\right>=\left< X_{0}^{\bar \sigma,\bar \sigma}\right>$)
reads:
\begin{equation} \label{eq:D}
\hat{D}(\vec{k},E) = \left[ \hat{G}_0^{-1}(E)
- \hat{P}(\vec{k},E) \hat{t}_{\vec{k}}
- \hat{P}(\vec{k},E) \hat{V}_{\vec{k}\vec{k}}^{\sigma,\sigma} \left< X_{0}^{\sigma,\sigma}\right>
+ \hat{\Sigma}(\vec{k},E) \right]^{-1} \hat{P}(\vec{k},E).
\end{equation}

Here, $\hat{G}_0^{-1}(E)$ is the exact local Green function,
$G_0^{\alpha \beta}(E) = \delta_{\alpha \beta} / \left[ {E - \left({\varepsilon_n - \varepsilon_{n'} } \right)} \right]$,
$\hat{\Sigma}(\vec{k},E)$ and $\hat{P}(\vec{k},E)$ are the self-energy and the strength operators, respectively.
The presence of the strength operator is due to the redistribution of the spectral weight between the Hubbard subbands, that is an intrinsic feature of the strongly correlated electron systems. It is also should be stressed that $\hat{\Sigma}(\vec{k},E)$ in Eq.~(\ref{eq:D}) is the self-energy in the $X$-operators representation and therefore it is different from the self-energy entering the Dyson equation for the weak coupling perturbation theory for $G_{\lambda}(\vec{k},E)$.

To calculate the strength operator $\hat{P}(\vec{k},E)$ we use the zero-loop approximation given by the replacement: $P^{\alpha \beta}(\vec{k},E) \to P^{\alpha \beta} = \delta_{\alpha \beta} F_\alpha $,
where $F_{\alpha(n,n')} = \left< X_f^{n,n} \right> + \left< X_f^{n',n'} \right>$ is the occupation factor.
Taking into account that in the considered paramagnetic phase
$\left< X_f^{\sigma,\sigma} \right> = \frac{1-x}{2}$, $\left< X_f^{0,0} \right> = x$,
after all substitutions and treating all $\vec{k}$-independent terms as the chemical potential renormalization, the generalized Dyson equation for the
Hamiltonian~(\ref{eq:H}) becomes:
\begin{equation}
\label{eq:D_H}
D(\vec{k},E) = \frac{(1+x)/2}{E - (\varepsilon_0 - \mu)
- \frac{1+x}{2} t_{\vec{k}} - \frac{1+x}{2} \frac{\tilde{t}_{\vec{k}}^2}{E_{ct}} \frac{1-x}{2}
+ \Sigma(\vec{k},E)}.
\end{equation}

To go beyond the Hubbard-I approximation we have to calculate $\Sigma(\vec{k},E)$. This was done in Ref.~\cite{korshunov2006} using an equations of motion method for the $X$-operators \cite{plakida1989}. The calculations resulted in the following expression:
\begin{eqnarray}
\label{eq:Sigma}
\Sigma(\vec{k}) &=& \frac{2}{1+x} \frac{1}{N} \sum\limits_{\vec{q}} \left\{
  \left[ t_{\vec{q}}
  - \frac{1-x}{2} J_{\vec{k}-\vec{q}}
  - x \frac{\tilde{t}_{\vec{q}}^2}{E_{ct}}
  - \frac{1+x}{2} \frac{2 \tilde{t}_{\vec{k}} \tilde{t}_{\vec{q}}}{E_{ct}} \right] K_{\vec{q}} \right.\nonumber\\
  &-& \left.
  \left[ t_{\vec{k}-\vec{q}}
  - \frac{1-x}{2} \left( J_{\vec{q}} - \frac{\tilde{t}_{\vec{k}-\vec{q}}^2}{E_{ct}} \right)
  - \frac{1+x}{2} \frac{2 \tilde{t}_k \tilde{t}_{\vec{k}-\vec{q}}}{E_{ct}}
  \right] \frac{3}{2} C_{\vec{q}} \right\}.
\end{eqnarray}
Here $N$ is the number of vectors in momentum space.
Also, the static spin-spin correlation function
\begin{equation}
\label{eq:C}
C_{\vec{q}} = \sum\limits_{f,g} e^{-i(f-g)\vec{q}} \left< X_f^{\sigma \bar\sigma} X_g^{\bar\sigma \sigma} \right> = 2 \sum\limits_{\vec{r}} e^{-i \vec{r} \vec{q}} \left< S_{\vec{r}}^z S_0^z \right>,
\end{equation}
and the kinematic correlation function
\begin{equation}
\label{eq:K}
K_{\vec{q}} = \sum\limits_{f,g} e^{-i(f-g){\vec{q}}} \left< X_f^{\sigma 0} X_g^{0\sigma} \right>,
\end{equation}
were introduced.

Kinematic correlation functions (\ref{eq:K}) are calculated straightforwardly
using the Green function~(\ref{eq:D_H}).
The spin-spin correlation functions for the $t-J$ model with the
three-site correlated hoppings $H_3$ were calculated in Ref.~\cite{valkov2005}
and the following expression for the Fourier transform of the spin-spin Green function was derived:
\begin{equation}
\label{eq:SpinGF}
\left<\left< X_{\vec{q}}^{\sigma \bar\sigma} \left| \right. X_{\vec{q}}^{\bar\sigma \sigma} \right>\right>_\omega
= \frac{A_{\vec{q}}(\omega)}{\omega^2 - \omega_{\vec{q}}^2},
\end{equation}
where $A_{\vec{q}}(\omega)$ and magnetic excitations spectrum $\omega_{\vec{q}}$ are given in Ref.~\cite{valkov2005} by the Eqs.~(25) and (26), respectively.

The following results were obtained by the self-consistent calculation of
the chemical potential $\mu$, the spin-spin correlation functions~(\ref{eq:C})
using Green function~(\ref{eq:SpinGF}), and the kinematic correlation
functions~(\ref{eq:K}) using Green function~(\ref{eq:D_H})
with the self-energy~(\ref{eq:Sigma}).

Parameters of the effective $t-t'-t''-J^*$ model were obtained directly from the {\it ab-initio} parameters of the multiband model, Table~\ref{table2}.
Their dependence on Ce concentration is presented in Table~\ref{table3}.
Note, here we took Cu-$4s$ orbitals into account implicitly through LDA+GTB method, as was described in Section~\ref{section:LDA}.
However, within the Hubbard operators technique it is possible to take these orbitals into account explicitly using more cumbersome methods like the one introduced in Ref.~\cite{Digor2006}.

\begin{table}
\caption {\label{table3} Doping-dependence of the effective $t-t'-t''-J^*$ model parameters (all values are in eV).
Note, in the notations of Ref.~\cite{mmk2005} the hopping matrix elements $t_{fg}$ corresponds to $-t_{fg}^{00}$, and $\tilde{t}_{fg} = -t_{fg}^{0S}$.
Also, $J_{fg} = 2 \tilde{t}_{fg}^{2} / E_{ct}$.
}
\begin{indented}
\item[] \begin{tabular}{lrrrrrr}
\br
Parameter        & $x=$0.00&  0.05&  0.10&  0.15&  0.20&  0.30\\
\mr
$-t \equiv -t_{01}$   & 0.552& 0.560& 0.561& 0.572& 0.572& 0.567\\
$-t' \equiv -t_{11}$  &-0.054&-0.053&-0.050&-0.056&-0.054&-0.052\\
$-t'' \equiv -t_{02}$ & 0.086& 0.087& 0.087& 0.070& 0.089& 0.088\\
$J \equiv J_{01}$     & 0.463& 0.477& 0.484& 0.488& 0.492& 0.486\\
$J' \equiv J_{11}$    & 0.007& 0.007& 0.007& 0.007& 0.007& 0.007\\
$J'' \equiv J_{02}$   & 0.012& 0.013& 0.013& 0.013& 0.013& 0.013\\
$-\tilde{t}_{01}$     & 0.680& 0.691& 0.695& 0.699& 0.701& 0.697\\
$-\tilde{t}_{11}$     &-0.085&-0.085&-0.081&-0.086&-0.084&-0.082\\
$-\tilde{t}_{02}$     & 0.111& 0.112& 0.112& 0.113& 0.113& 0.112\\
\br
\end{tabular}
\end{indented}
\end{table}

\begin{figure}
\centering
\includegraphics[width=0.45\linewidth,angle=270]{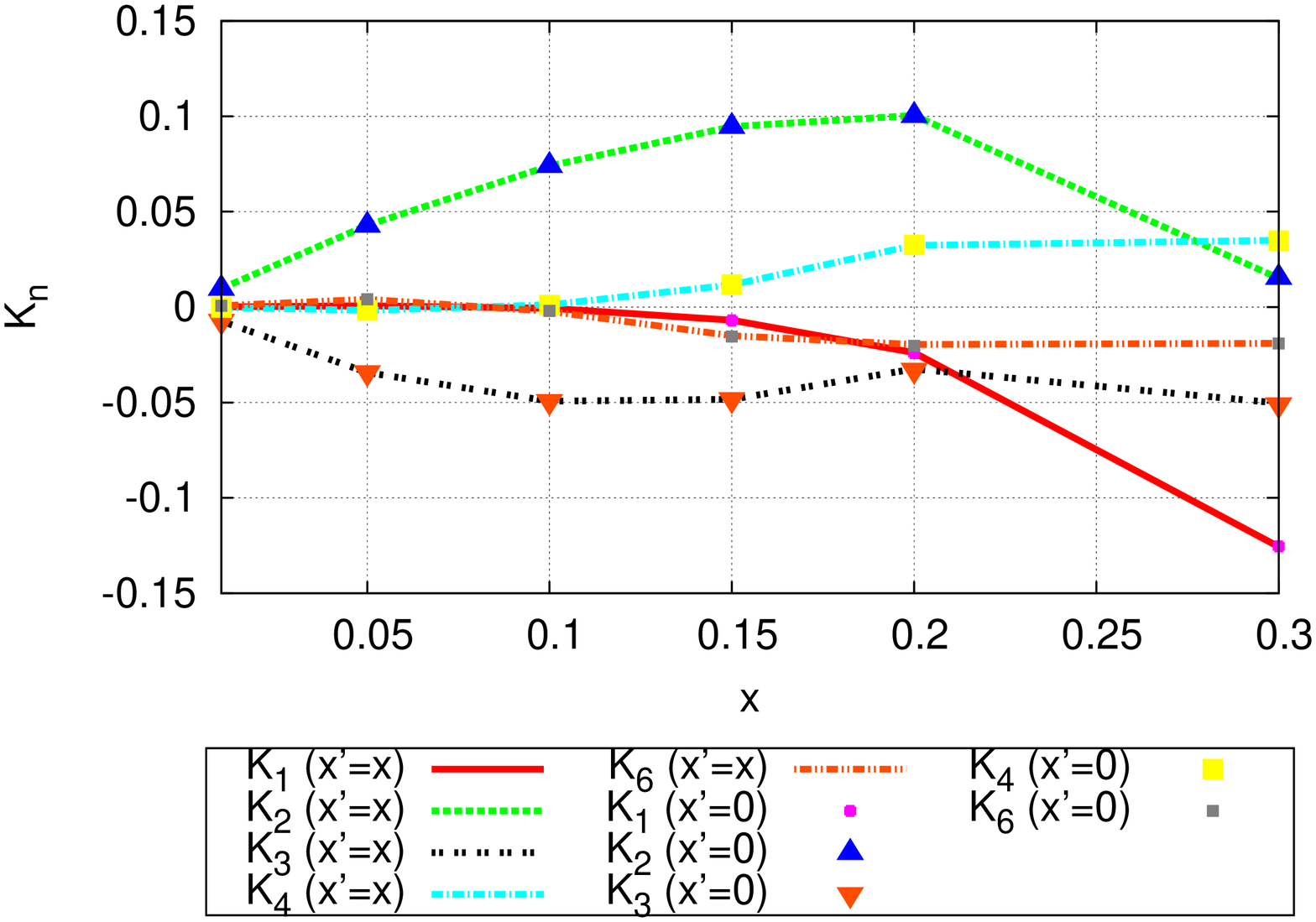}
\includegraphics[width=0.45\linewidth,angle=270]{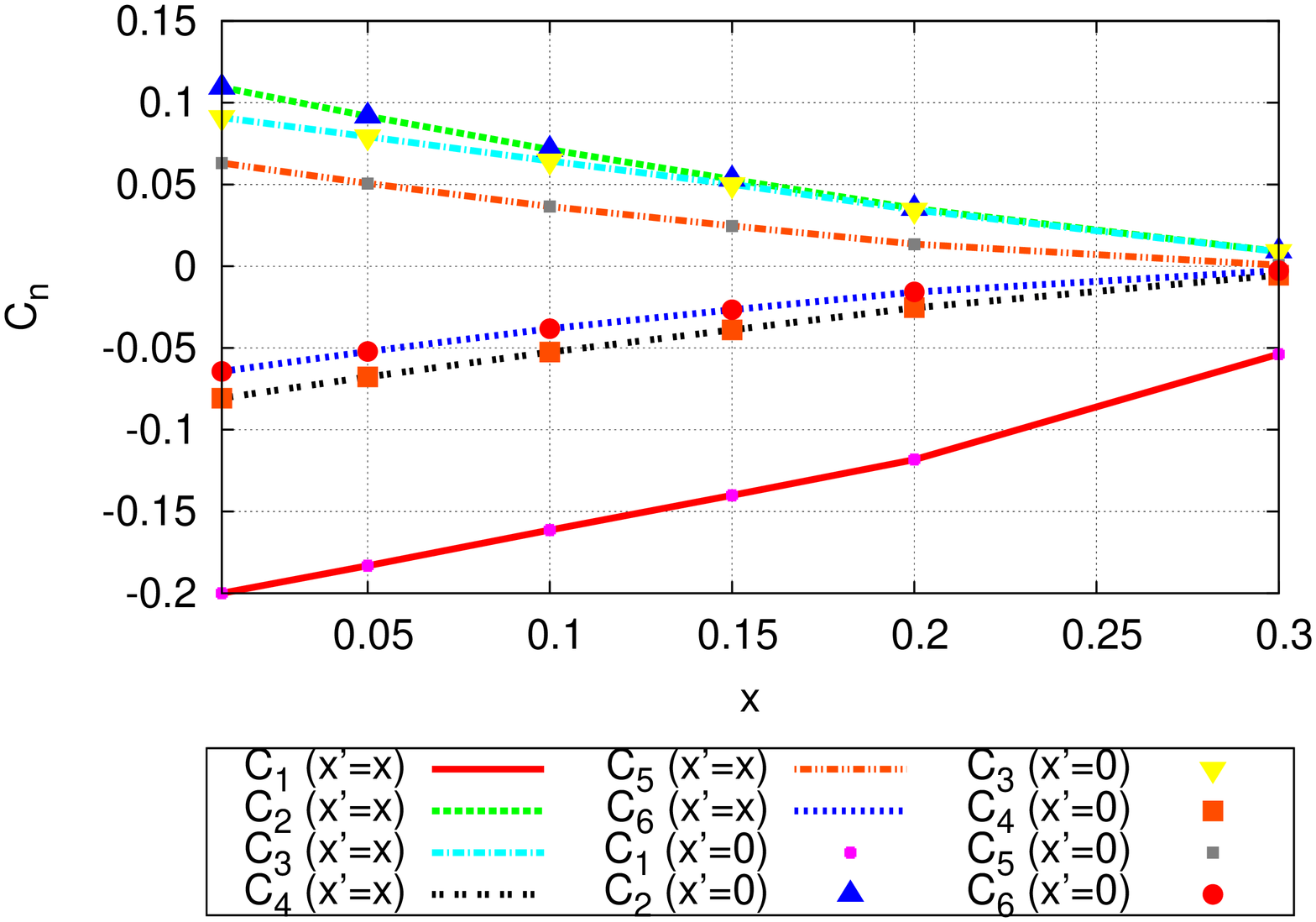}
\caption{(color online) Doping dependent evolution of the kinematic (upper panel) and the spin-spin (lower panel) correlation functions within the $t-t'-t''-J^*$ model. Index $n$ enumerates the real space vectors connecting the neighboring sites: $n=1$ for the nearest-neighbors, $n=2$ for the next nearest neighbors, and so on.}
\label{fig:tJ3modelKC}
\end {figure}

Our results for the doping dependence of the kinematic and spin-spin correlation functions are shown in Fig.~\ref{fig:tJ3modelKC}.
Here, variable $x'$ denotes the concentrations for which the one-electron parameters were calculated. Thus, $x'=0$ corresponds to the absence of the one-electron mechanism of doping dependence, while $x'=x$ corresponds to the presence of this mechanism. However, the many-body mechanism is present in both cases.
Note, the behavior of all correlation functions is almost identical for the cases of presence and absence of one-electron mechanism of doping dependence.
Also note, the kinematic correlation functions, $K_n$, possess a very nontrivial doping dependence. For low concentrations, $x<0.2$, due to the strong magnetic correlations the hoppings to the nearest neighbors are suppressed leading to the small value of $K_1$, while $K_2$ and $K_3$ are not suppressed. Upon increase of the doping concentration above $x \approx 0.2$, magnetic correlations decrease considerably and the nearest-neighbor kinematic correlation function $K_1$ increase, reviving the almost Fermi liquid behavior: $K_1$ becomes largest of all $K_n$'s, while the magnetic correlation functions, $C_n$, are strongly suppressed.

\begin{figure}
\centering
\includegraphics[width=0.75\linewidth]{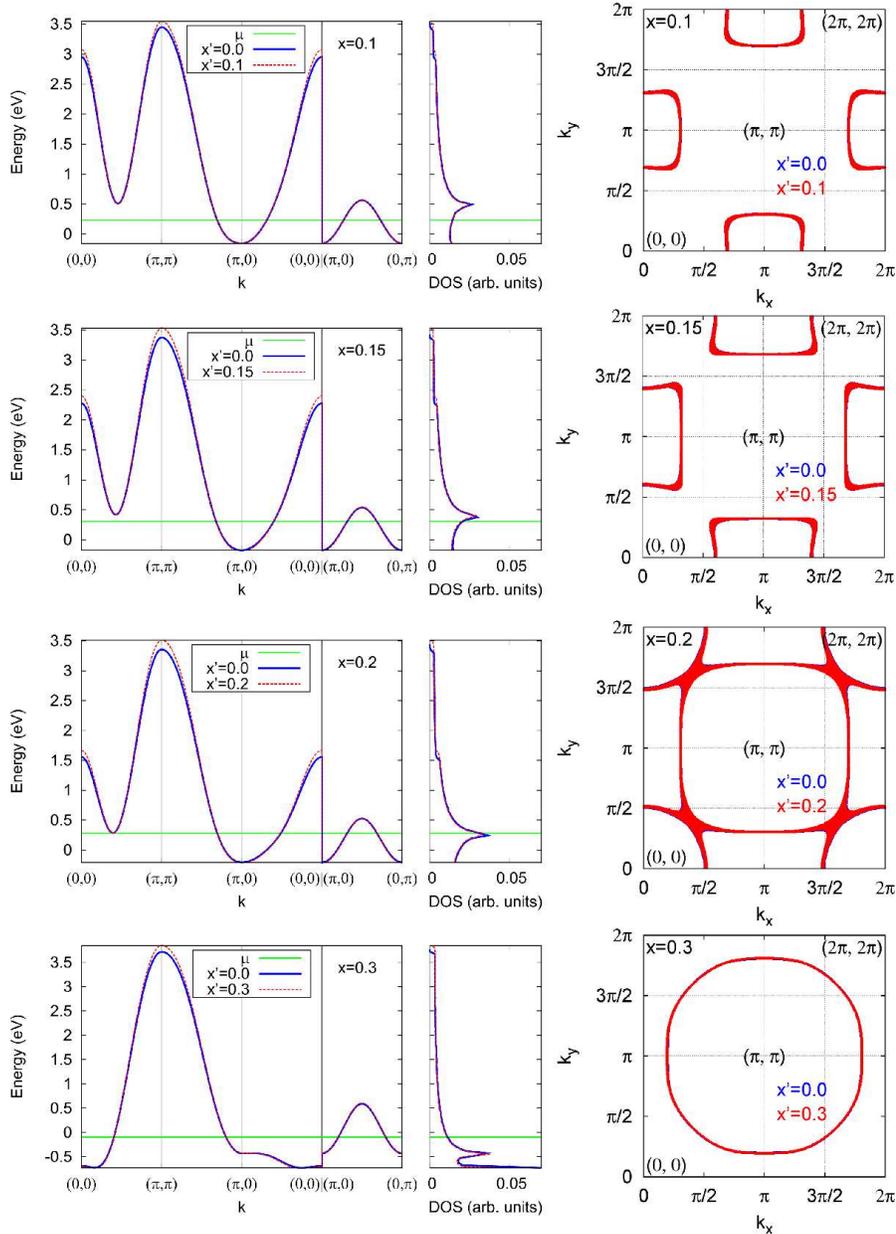}
\caption{Doping-dependent quasiparticle dispersion (on the left) and
Fermi surface (on the right) in the spin-liquid phase of n-type cuprates. The position of the chemical potential is denoted by the horizontal (green) line. The solid blue curves correspond to the calculations without taking the one-electron mechanism of concentration dependence into account ($x'=0$ case). The dashed red curves represent results of the calculations in which both many-body and one-electron mechanisms were considered ($x'=x$ case). Note, the Fermi surfaces for both $x'=0$ and $x=x$ cases are almost indistinguishable.}
\label{fig:tJ3modelEk}
\end {figure}

So, we can clearly define one point of the crossover, namely, $x_m \approx 0.2$. The system behavior is quite different on the different sides of these point, although there is no phase transition with symmetry breaking occurs.
Apparently, this crossover is closely connected to the change of
the Fermi surface topology with doping. Fermi surface evolution together with
the quasiparticle dispersion is shown in Fig.~\ref{fig:tJ3modelEk}.
For small $x$, the electron pockets around $(\pm \pi,0)$ and $(0, \pm \pi)$
points are present at the Fermi surface. Upon increase of the doping concentration these pockets become larger and merge together at $x=0.2$. For higher concentrations, the Fermi surface appears to be a large hole-like one, shrinking toward the $(\pi, \pi)$ point. Thus, the topology of the Fermi surface changes at the same doping $x_m$, where the point of crossover is situated. For the first time the ``electronic transition'' accompanying the change of the Fermi surface topology, or the so-called Lifshitz transition, was described in Ref.~\cite{lifshitz1960}. Now such transitions referred as a quantum phase transitions with a co-dimension$=1$ (see e.g. paper~\cite{volovik2006}).

Note, when the Fermi surface topology changes at a quantum critical concentration $x_m=0.2$ the density of states at the Fermi level also exhibit significant transformations. This results in the different behavior of the kinematic and magnetic correlation functions on the different sides of this crossover point. And, of course, the changes in the density of states at the Fermi level will also result in the significant changes of such observable physical quantities as the resistivity and the specific heat.

Above the critical concentration, the Van Hove singularity in DOS is due
to the flat dispersion around the $(\pi, 0)$ point. However, for $x < x_m$ this singularity is due to the states near of the $(\pi/2, \pi/2)$ point. This is the result of the many-body interactions that can not be obtained within the
one-electron band theory.

Concerning the role of the short range magnetic order and three-site hopping terms in the n-type cuprates we would like to stress that due to the scattering on the magnetic excitations with the AFM wave vector $\vec{Q}=(\pi, \pi)$
the states near the $(\pi, \pi)$ point are pushed above the Fermi level,
and the local symmetry around the $(\pi/2, \pi/2)$ points is restored for
low doping concentrations (see Fig.~\ref{fig:tJ3modelEk}).
In other words, the short range magnetic order ``tries'' to restore
the AFM symmetry around $(\pi/2, \pi/2)$ point. In our calculations,
the short range magnetic fluctuations are taken into account via
the spin-spin correlation functions~(\ref{eq:C}).

Now we proceed to the comparison of the one-electron and many-body
mechanisms of the doping dependence. In Fig.~\ref{fig:tJ3modelEk} the quasiparticle dispersion without one-electron mechanism is shown by the solid blue curves. Apparently, its difference from the case when both one-electron and many-body mechanisms are present is negligibly small. In the latter case the bandwidth is slightly renormalized while band dispersion retains the same character. Moreover, the Fermi surfaces for both cases are very similar and the quantum phase transition will be at the same concentration, $x_m=0.2$.

\section{Conclusion}
In the present work we report the combined investigation of the one-electron
and the many-body mechanisms of the electronic structure doping dependence for the High-T$_c$ compound Nd$_{2-x}$Ce$_x$CuO$_4$. The electronic structure
calculations were performed within the hybrid LDA+GTB scheme. For the both
antiferromagnetic and paramagnetic spin-liquid phases we demonstrate
that the main effect on the electronic structure is provided by the many-body
mechanism, whereas the one-electron contribution leads to a rather small
quantitative modifications which do not change the picture qualitatively.
The role of the many-body mechanism is very important because of the strong
electronic correlations present in the underdoped cuprates.

\ack
This work was supported in part by RFBR Grants 06-02-16100 (M.M.K., V.A.G., S.G.O.),
06-02-90537-BNTS (M.M.K., V.A.G., S.G.O., I.A.N., Z.V.P.),
05-02-16301 (I.N.), 05-02-17244 (I.N.),
by the joint UrO-SO project 74, and programs of the Presidium of the Russian
Academy of Sciences (RAS) ``Quantum macrophysics'' and of the
Division of Physical Sciences of RAS ``Strongly correlated electrons
in semiconductors, metals, superconductors and magnetic materials''.
M.M.K. acknowledge support form INTAS (YS Grant 05-109-4891).
I.N. and Z.P. acknowledge support from the Dynasty Foundation and
International Center for Fundamental Physics in Moscow program for
young scientists.
I.N. appreciate the support from the grant of President of Russian
Federation for young PhD MK-2242.2007.02.
Z.V. is supported by UrO grant for young scientists.

\end{document}